\documentclass{ws-ijmpa}
\renewcommand{\bi}{\bibitem}
\newcommand{\be}{\begin{eqnarray}}
\newcommand{\ee}{\end{eqnarray}}
\newcommand{\bea}{\begin{array}}
\newcommand{\eea}{\end{array}}
\newcommand{\nn}{\nonumber}
\newcommand{\GeV}{\mbox{GeV}}

\begin{document}

\markboth{Yuji Kajiyama}
{R-Parity Violation and Family Symmetry}

%
\catchline{}{}{}{}{}
%

\title{R-Parity Violation and Family Symmetry
\footnote{To appear in the proceedings of CTP symposium on Supersymmetry at LHC: 
Theoretical and Experimental Perspectives, The British University in Egypt, Cairo, Egypt, 11-14 March 2007. }
}

\author{YUJI KAJIYAMA}

\address{National Institute of Chemical Physics and Biophysics,\\
Ravala 10, Tallinn 10143, Estonia\\
yuji.kajiyama@kbfi.ee}

\maketitle

\begin{history}
\received{15 June 2007}
\end{history}

\begin{abstract}
In this talk, we investigate the implications of R-parity violating (RPV) operators in a model 
with family symmetry \cite{kaji}.
Family symmetry can determine the form of RPV operators 
as well as the Yukawa matrices.  
We consider a concrete model with non-abelian discrete 
symmetry $Q_6$, which has only three RPV trilinear operators with no baryon number violating terms. 
We find that ratios of decay rates of the lepton flavor violating processes 
are fixed thanks to the family symmetry, predicting $BR(\tau \to 3e)/BR(\tau \to 3\mu)
\sim 4 m_{\mu}^2/m_{\tau}^2$.

\keywords{Supersymmetric models; Flavor symmetries; Masses and mixing.}
\end{abstract}

\ccode{PACS numbers: 12.60.Jv,~11.30.Hv,~12.15.Ff,~14.60.Pq}

\section{Introduction}
Despite the remarkable success of the gauge sector of the Standard Model (SM), there still exist some problems in the Higgs and Yukawa sectors. The Yukawa matrices are responsible for the masses and mixings of matter fermions: quarks and leptons. The Yukawa sector in the SM can give experimentally consistent masses and mixings, because it contains more free parameters than the number of observables in general. 
There is no predictivity in the Yukawa sector because of this redundancy of the parameters.
One of the ideas to overcome this issue is to introduce a family symmetry (flavor symmetry), which is the 
symmetry between generations. In this paper we consider a concrete model which is symmetric under the binary dihedral group $Q_6$ \cite{babukubo,kajiyama}. 
  
On the other hand, in the Higgs sector, the most important problem is that the Higgs boson has not been experimentally discovered yet. 
Discovery of the Higgs boson is expected at the Large Hadron Collider (LHC). 
In the SM, the Higgs mass 
is quadratic divergent. This problem is solved by introducing Supersymmetry (SUSY) 
at $O(1)$ TeV. 
The Minimal Supersymmetric Standard Model (MSSM) has low energy SUSY. 
In general it contains gauge symmetric, lepton and baryon number violating operators 
\be
W_{\not R}=\frac 12 \lambda_{ijk}L_i L_j E^c_k 
+\lambda'_{ijk}L_i Q_j D^c_k
+\frac 12 \lambda''_{ijk}U^c_i D^c_j D^c_k
+\mu_i H_u L_i
\label{rvmssm}
\ee
in addition to the usual Yukawa couplings and $\mu$-term. The asymmetric properties $\lambda_{ijk}=-\lambda_{jik}$ and $\lambda''_{ijk}=-\lambda''_{ikj}$ mean that $9+27+9+3=48$ (complex) parameters are included in this interactions. These couplings generate unacceptable processes such as   Lepton Flavor Violating (LFV) processes and proton decay. The conservation of R-parity 
\cite{farrar,rpreview} 
\be
R=(-1)^{3B+L+2s},
\ee
where $B$, $L$ and $s$ denote baryon, lepton number and spin of the particles, respectively, is one possibility to forbid the couplings Eq.(\ref{rvmssm}). From the definition, R-parity is $+1$ for all SM particles and $-1$ for all their superpartners. 
However, R-parity is not the only possible choice to forbid the interactions. 
Matter- or lepton- and baryon-parity \cite{ibanez} can be also a possibility. 
On the other hand, without R-parity,  these coupling constants have to be 
strictly constrained not to conflict with experimental data. 
Constraints on the R-parity violating couplings have been obtained by many authors 
from LFV processes \cite{hall,lfv,hinchliffe,gouvea,kim,chaichian}, neutrino mass 
\cite{hall,rnumass,joshipura,banks,bisset}, neutral meson system 
\cite{mesonmixing,carlos,bhattacharyya,saha,kundu,wang,grossman,jang,dreiner}, 
proton decay \cite{hinchliffe,benhamo,pdecay}, and so on \cite{rpreview,choudhury,allanach}. 

Family symmetries also constrain the form of R-parity interactions \cite{banks,benhamo,abelian,nonabelian} as well as the Yukawa matrices. 
In the model that we consider \cite{babukubo,kajiyama}, the $Q_6$ family symmetry reduces the 45 trilinear couplings to three: 
$\lambda$, $\lambda'_{1}$ and $\lambda'_{2}$. The baryon number violating couplings $\lambda''_{ijk}U^c_i D^c_j D^c_k$ are forbidden by the symmetry in our model, so it is guaranteed by the symmetry that the R-parity violating operators do not induce proton decay. 

In this paper, we study the phenomenology of  the three R-parity violating interactions in the model with 
$Q_6$ family symmetry \cite{babukubo,kajiyama}. First, we obtain upper bounds on three coupling constants 
$\lambda$ and $\lambda'_{1,2}$ from the experimental constraints. 
Next we focus on LFV processes induced by $\lambda$.  
The $\lambda LLE^c$ coupling generates the LFV decays 
$\ell_m^- \to \ell_i^- \ell_j^- \ell_k^+(m,i,j,k$ denote the flavor of the charged lepton) at tree level, and 
the Branching Ratios ($BR$) of the decay processes are proportional to $\lambda^4$.
Therefore, the ratios of these processes are independent of $\lambda$ and can be predicted 
unambiguously to be  
$BR(\tau \to eee)/BR(\tau \to \mu\mu\mu)\sim~4m_{\mu}^2/m_{\tau}^2$. 
It reflects the properties of the family symmetry. 
We introduce the $Q_6$ symmetric model in the next section, and derive the predictions in the Sec. 3. 
Sec. 4 is conclusions.

This talk is based on ref.\cite{kaji}.
\section{The Model}
\subsection{Group Theory of $Q_6$ and Assignment}
The binary dihedral group $Q_N(N=2,4,6,...)$ is a finite subgroup of $SU(2)$ and defined by the following set of $2N$ elements
\be
Q_N&=&\left\{ 1,A,A^2,...,A^{N-1},B,AB,...,A^{N-1}B\right\} ,
\ee
where two dimensional representation of matrix $A$ and $B$ is given by
\be
A&=&\left(\bea{cc} 
\cos \phi_N & \sin \phi_N \\
-\sin \phi_N & \cos \phi_N \\
\eea \right),~\phi_N=\frac{2 \pi}{N},~~~
B=\left( \bea{cc} 
i & \\
 & -i \\
\eea \right).
\label{QNdef}
\ee
In this note, we consider the case of $N=6$ {\footnote{In $Q_8$ model of ref\cite{q8}, 
 definition of $Q_N$ is different from ours. Our $Q_N$ is equivalent to their $Q_{2N}$. }}.

$Q_6$ group contains 2 two-dimensional irreducible representations (irreps), 
${\bf 2}_1,{\bf 2}_2$ and 4 one-dimensional ones ${\bf 1}_{+,0},{\bf 1}_{+,2},{\bf 1}_{-,1},{\bf 1}_{-,3}$, where ${\bf 2}_1$ is pseudo real and ${\bf 2}_2$ is real representation. 
In the notation of ${\bf 1}_{\pm,n}(n=0,1,2,3)$, $\pm$ stands for the change of sign under the transformation by matrix $A$, and $n$ the factor $\exp(i n \pi/2)$ by $B$. So ${\bf 1}_{+,0}$ and ${\bf 1}_{+,2}$ are 
real representations, while ${\bf 1}_{-,1}$ and ${\bf 1}_{-,3}$ are complex conjugate to each other. 
We consider a extension of Supersymmetric Standard Model with $Q_6$ family symmetry, 
where three generations of matter and Higgs fields are assumed to be embedded into 
two- and one- dimensional irreps. of $Q_6$ group.   
Assignment of $Q_6$ group for the quark, 
lepton and Higgs chiral supermultiplets are shown below in an   
obvious notation:
 \be
{\bf 2}_1~~&:&~~Q_I, \nn \\
{\bf 2}_2~~&:&~~U_I^c,~D_I^c,~\hat L_I,~E_I^c,~N_I^c,~H^u_I,~\hat H^d_I,\nn \\
{\bf 1}_{+,0}~~&:&~~ L_3,~E_3^c,\nn \\
{\bf 1}_{+,2}~~&:&~~Q_3,~Y, \label{assignment2} \\
{\bf 1}_{-,1}~~&:&~~U_3^c,~D_3^c,~H^u_3,~H^d_3, \nn \\
{\bf 1}_{-,3}~~&:&~~N_3^c. \nn
\ee
The generation indices $I,J,...=(1,2)$ are applied to the $Q_6$ doublet, and 
$i,j,...=(1,2,3)$ to three generations throughout the paper. 
$Y$ is gauge singlet Higgs supermultiplet to give neutrino mass by seesaw mechanism.
In a model without R-parity conservation, there is no distinction between lepton doublet 
and down type Higgs doublet, because both have the same gauge quantum numbers. 
In our model, $L_3$ and $H_3^d$ are distinguishable because they have different $Q_6$ quantum 
number to each other, although $L_I$ and $H_I^d$ belong to the same irreps. 
Therefore,   
we have written these fields as $\hat L_I,\hat H^d_I$, 
and physical lepton doublet and 
down type Higgs doublet ($ L_I,H^d_I$) should be written as linear combination of these.

\subsection{Fermion mass matrices and 
diagonalization}
 We assume that physical Higgs fields acquire complex vacuum expectation values (VEVs) 
$\langle H^{u,d}_I \rangle=v^{u,d}_D e^{i \theta^{u,d}}/2$ and 
$\langle H^{u,d}_3 \rangle=v^{u,d}_3 e^{i \theta^{u,d}_3}/\sqrt{2}$ in order to avoid SUSY CP problem in soft SUSY breaking sector.  
 
 The quark mass matrices are given by

\be
{\bf {m}}^u = m_t\left(\begin{array}{ccc}
0 & q_u/y_u & 0  \\ -q_u/y_u & 0 & b_u\\
0 &   b_u'  &  y^2_u \\
\end{array}\right),
\label{muhat}
\ee
and similarly for ${\bf m}^d$, where phases from VEVs have been absorbed into 
a part of the CKM matrix.  

Many set of the parameters give observables consistent with experimental data, and 
one example is
 \be
\theta_q & = &\theta^d_3-\theta^d-\theta^u_3
+\theta^u  =-1.25,
q_u=0.0002150,
b_u=0.04440,
b'_u=0.09300,\nn\\
y_u &=&0.99741,
q_d =0.005040,
b_d=0.02500,
b'_d=0.7781,
y_d=0.7970.
\label{parameters}
\ee
One can easily obtain unitary matrices in explicit form 
which diagonalize mass matrices from these parameters.  
Moreover, since the CKM parameters and the quark masses are related to 
each other because of the 
family symmetry, we find that {\em nine} independent parameters (Eq.(\ref{parameters})) of the model 
can well describe {\em ten} physical observables: there is {\em one} prediction. 
An example of the prediction is $|V_{td}/V_{ts}|$,
whose experimental value has been obtained from
the measurement
of  the mass difference $\Delta m_{B_s}$ of the $B_s^0$ meson \cite{deltams}:
\be
\mbox{Model}&:& 
|V_{td}/V_{ts}| =0.21-0.23,\nn\\
\mbox{Exp.} &:& |V_{td}/V_{ts}|
=0.208\begin{array}{c}+0.001\\-0.002\end{array}
\mbox{(exp.)}
\begin{array}{c}+0.008\\-0.006\end{array}
\mbox{(theo.)}.
\label{pred3}
\ee

The mass matrix in the charged lepton sector is:
\be
{\bf m}^{e} &=&\frac{1}{2} \left(\begin{array}{ccc}
- Y_{c}^{e}& 
Y_{c}^{e}
&  Y_b^{e} \\
 Y_{c}^{e} & 
  Y_{c}^{e} &
 Y_b^{e}  \\
Y_{b'}^{e}  & 
Y_{b'}^{e} & 
0 \\
\end{array}\right)  v_D^d e^{-i \theta^d}.
\label{me}
\ee
One finds that  $U_{eL}$ and $ U_{eR}$ can be approximately 
written
as
\be
U_{eL}&\simeq&\left( \bea{ccc} 
\epsilon_e & 1/\sqrt{2}
&1/\sqrt{2} \\
-\epsilon_e & -1/\sqrt{2}
&1/\sqrt{2} \\
1 & -\sqrt{2} \epsilon_e &0\\
\eea \right),~ 
U_{eR}\simeq \left( \bea{ccc} 
0&-1&0 \\
1 &0&\epsilon_{\mu} \\
-\epsilon_{\mu} &0 & 1 \\
\eea \right)e^{i \theta^d}
, \label{UeLR}
\ee
where terms of ${\cal O}(\epsilon^2)$ are neglected, 
and small parameters $\epsilon_e,\epsilon_{\mu}$ are defined as
\be
\epsilon_e=\frac{m_e}{\sqrt{2}m_{\mu}}=3.42 \times 10^{-3},~
\epsilon_{\mu}=\frac{m_{\mu}}{m_{\tau}}=5.94 \times 10^{-2}.
\label{epsilon}
\ee
The upper-right $2 \times 2$ block of $U_{eL}$ is the origin of 
maximal mixing of the atmospheric neutrino oscillation.  

As for the neutrino sector, we assume that
a see-saw mechanism \cite{seesaw} takes place.
However, we do not present the details of the neutrino sector here because 
there is no need to know it in the following analysis. We obtain some specific predictions of our model:
(i) only an inverted mass hierarchy $m_{\nu_3}<m_{\nu_1},m_{\nu_2}$ is consistent with 
the experimental constraint $|\Delta m_{21}^2|< |\Delta m_{23}^2|$, 
(ii) the $(e,3)$ element of the MNS matrix is given by $|U_{e3}| \simeq \epsilon_e$. 
See Refs. \cite{kajiyama,kubo} for details.

\section{R-Parity Violation}
Since $Q_6$ family symmetry controls the whole flavor structure of the model, the form of the R-parity violating couplings are also constrained by the family symmetry. We find that only possible trilinear couplings 
allowed by the symmetry can be written as
\be
W_{\not R}=\lambda L_3 L_I E_I^c+\lambda_1'  L_I (i \sigma^2)_{IJ}Q_3 D_J^c
+\lambda_2'  L_I (\sigma^1)_{IJ}Q_J D_3^c
\label{rparity}
\ee
in the physical lepton doublet $L_I$. 
Here the superpotential $W_{\not R}$ is written in the flavor eigenstates, so 
the unitary matrices $U_{u(L,R)},~U_{d(L,R)}$ and $U_{e(L,R)}$  
should appear when we rotate the fermion components into their mass eigenstates.
On the other hand, these matrices do not appear from sfermion components, because 
we approximate that sfermions are in the mass eigenstate basis. This approximation is 
valid in our model, because scalar masses in the soft SUSY breaking sector have diagonal form 
because of non-Abelian property of the family symmetry.   
In the present model, there are only three R-parity violating trilinear interactions allowed by the 
family symmetry, and baryon number violating terms $\lambda''_{ijk}U^c_i D^c_j D^c_k$ 
are forbidden by the symmetry.  
It should be compared with the MSSM case in which there are 
45 trilinear couplings. 
These interactions can generate a lot of new processes which have not been observed yet 
such as lepton flavor violating (LFV) processes, 
or new contributions to already observed processes.
Many authors have studied phenomenology of R-parity violation and  
obtained constraints on each coupling constant corresponding to each process 
in the MSSM case.
\footnote{See Ref. \cite{rpreview,allanach} and references therein.}

In this section, we obtain constraints on the coupling constants $\lambda, \lambda'_{1,2}$ 
at the weak scale.
Since the various new processes generated by the interactions
$W_{R\hspace{-2mm}/}$ depend only on the three coupling constants, we can predict ratios 
of new processes independent of $\lambda$s.  
We will also find the ratios of the LFV processes in this section.
In the following analysis, we assume that the R-parity violating couplings $\lambda$ and $\lambda'_{1,2}$ are real and positive \footnote{CP violation induced by the R-parity violating trilinear couplings 
in the soft SUSY breaking sector has been studied in Ref. \cite{abel}.}.

\subsection{Constraint on $\lambda$}
\begin{figure}[t]
\unitlength=1mm
\hspace{0.5cm}
\begin{picture}(60,30)
\includegraphics[width=5cm]{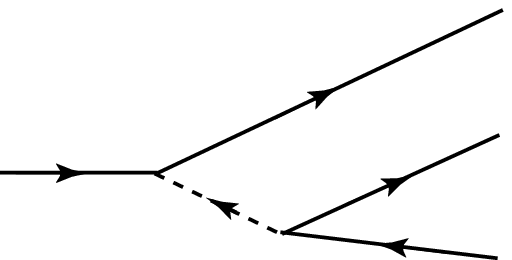}
\put(-37,11){$\lambda$}
\put(-50,11){$\ell_m^-$}
\put(1,12){$\ell^-_j$}
\put(1,25){$\ell^-_i$}
\put(1,-2){$\ell^+_k$}
\put(-32,2){$\tilde \nu$}
\put(-24,-2){$\lambda$}
\hspace{1.5cm}
\includegraphics[width=5cm]{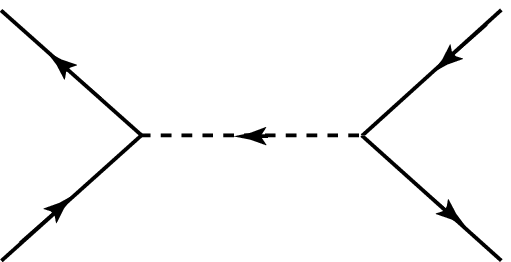}
\put(-26,15){$\tilde \nu$}
\put(-36,15){$\lambda'$}
\put(-17,15){$\lambda'$}
\put(-4,-4){$d_j$}
\put(-4,27){$\bar d_i$}
\put(-49,-4){$d_i$}
\put(-49,27){$\bar d_j $}
\end{picture}
\vspace{1.0cm}
\caption{Contributions to the decay processes 
$\ell_m^- \to \ell_i^- \ell_j^- \ell_k^+$ by coupling $\lambda$ (left) 
and the neutral meson mixing by $\lambda'$ (right).}
\label{lle}
\end{figure}

In this subsection, we consider the constraint on $\lambda L_3 L_I E^c_I$ operator.
The most stringent constraint on $\lambda$ is obtained from both $\mu \to eee$ 
and neutrinoless double beta decay, both the processes give similar bound. 
Here we explicitly show only $\mu \to eee$ process\cite{hall,lfv,hinchliffe,gouvea,choudhury}.
The decay process $\ell_m^- \to \ell_i^- \ell_j^- \ell_k^+$ is generated by tree-level 
$t$- and $u$- channel sneutrino 
exchange (Fig. \ref{lle}), and its effective Lagrangian is given by
\be
{\cal L}_{eff}=\lambda^2 A_L \left(\bar \ell_iP_L \ell_m\right) \left( \bar \ell_jP_R \ell_k\right)
+\lambda^2 A_R\left(\bar \ell_iP_R \ell_m\right) \left( \bar \ell_jP_L \ell_k\right)+(i \leftrightarrow j),
\label{mueee}
\ee
where the coefficients are 
\be
A_L&=& \frac{1}{m_{\tilde \ell 1L}^2}(U_{eR}^\dag)_{iJ}(U_{eR})_{Jk}
(U_{eL}^\dag)_{j3}(U_{eL})_{3m}\nn \\
&~&+\frac{1}{m_{\tilde \ell 3L}^2}(U_{eR}^\dag)_{iJ}(U_{eL})_{Jm}
(U_{eL}^\dag)_{jK}(U_{eR})_{Kk},\\
A_R&=&\frac{1}{m_{\tilde \ell 1L}^2}(U_{eR}^\dag)_{jJ}(U_{eR})_{Jm}
(U_{eL}^\dag)_{i3}(U_{eL})_{3k}\nn \\
&~&+\frac{1}{m_{\tilde \ell 3L}^2}(U_{eR}^\dag)_{jJ}(U_{eL})_{Jk}
(U_{eL}^\dag)_{iK}(U_{eR})_{Km},
\label{mueeeamp}
\ee
with mixing matrices in Eq.(\ref{UeLR}). 
In our approximation, sneutrino mass is the same as that of the left-handed slepton.
From this Lagrangian, the branching ratio of $\mu \to eee$ is given by, 
\be
BR(\mu \to eee)=\frac{4 \lambda^4}{64 G_F^2}\left[ |A_L|^2+|A_R|^2\right]
BR(\mu \to e \bar \nu_e \nu_{\mu}).
\label{mueeebr}
\ee
The requirement that this branching ratio should not exceed the experimental bound 
$BR(\mu\to eee)^{exp}<1.0 \times 10^{-12}$ provides a constraint on $\lambda$
\be
\lambda<1.4 \times 10^{-2} \left( \frac{m_{\tilde \ell L}}{100 \GeV}\right),
\label{lam1}
\ee
where we have assumed $m_{\tilde \ell 1L}=m_{\tilde \ell 3L}\equiv m_{\tilde \ell L}$ 
in order to forbid the contribution to FCNC processes from the soft scalar mass terms.

\subsection{Constraints on $\lambda_{1}'$ and $\lambda_{2}'$}
Constraints on $\lambda_{1,2}'$ are obtained from neutral meson mixings 
\cite{mesonmixing,carlos,bhattacharyya,saha,kundu,wang,choudhury}, and 
on the products $\lambda \lambda_{1}'$ and $\lambda \lambda_{2}'$ 
from leptonic decays of neutral mesons 
\cite{bhattacharyya,grossman,jang,dreiner,choudhury}.
Since both processes are generated at tree level, these give the most stringent bounds on 
$\lambda'_{1,2}$. Although $\mu -e$ conversion in nuclei \cite{gouvea,kim} is also generated at tree level by $\lambda'_{1,2}$, bounds from this process are weaker than those 
from neutral meson system. 

The neutral meson mixing is generated at the tree level through the exchange of a sneutrino 
in both $s$- and $t$- channels (Fig. \ref{lle}).
For $K^0-\bar K^0$ mixing, the effective Hamiltonian is obtained as
\be
{\cal H}_{eff}=\frac{\Lambda'_{I21}\Lambda'^{*}_{I12}}{m^2_{\tilde \ell 1L}}
(\bar d_R s_L)(\bar d_L s_R),
\label{k0mixham}
\ee
where
\be
\Lambda'_{Ijk}=\lambda'_1(U_{dR}^\dag)_{jJ}(U_{dL})_{3k}(i \sigma^2)_{IJ}
+\lambda'_2 (U_{dR}^\dag)_{j3}(U_{dL})_{Jk}(\sigma^1)_{IJ}.
\label{k0mixamp}
\ee
We require that these additional contributions to the mass difference of neutral K meson are 
smaller than its experimental value:
The assumptions $\theta^d_1-\theta^d_3=0$ and $\lambda'_1=\lambda'_2$
lead to the most stringent constraints on $\lambda'_{1,2}$:
\be
\lambda'_1=\lambda'_2<3.1 \times 10^{-3}\left( \frac{m_{\tilde \ell L}}{100 \GeV}\right).
\label{lamp12}
\ee

From leptonic decays of neutral Kaon, we get constraints on $\lambda \lambda'_{1,2}$, which are   
\be
\lambda \lambda'_1 &<& 5.4 \times 10^{-7} \left( \frac{m_{\tilde \ell L}}{100 \GeV}\right)^2
 \label{lamlamp1}
\ee
from $K_L \to \mu^{\mp}e^{\pm}$, and 
\be
\lambda \lambda'_2 &<& 1.1 \times 10^{-8} \left( \frac{m_{\tilde \ell L}}{100 \GeV}\right)^2
\label{lamlamp2}
\ee
from $K_L \to e^- e^+$.

\subsection{Predictions for Lepton Flavor Violating processes}
\begin{figure}[t]
\unitlength=1mm
\begin{picture}(60,31)
\hspace{0.7cm}
\includegraphics[width=5cm]{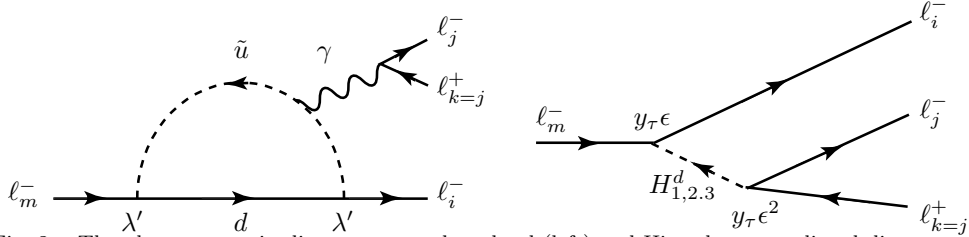}
\put(-56,1){$\ell^-_m$}
\put(1,15){$\ell_{k=j}^+$}
\put(1,23){$\ell_j^-$}
\put(1,1){$\ell^-_i$}
\put(-15,20){$\gamma$}
\put(-26,20){$\tilde u$}
\put(-26,-3){$d$}
\put(-41,-3){$\lambda'$}
\put(-13,-3){$\lambda'$}
\hspace{1.3cm}
\includegraphics[width=5cm]{mutoeee.eps}
\put(-37,11){$y_{\tau} \epsilon$}
\put(-50,11){$\ell_m^-$}
\put(1,12){$\ell^-_j$}
\put(1,25){$\ell^-_i$}
\put(1,-2){$\ell^+_{k=j}$}
\put(-35,2){$H^d_{1,2.3}$}
\put(-24,-2){$y_{\tau}\epsilon^2$}
\end{picture}
\caption{The photon penguin diagram at one loop level (left) and 
Higgs boson mediated diagram at tree level (right) to the decays 
$\ell_m^- \to \ell_i^- \ell_j^- \ell_{k=j}^+$. These contributions can be negligible 
compared to the tree level processes induced by the coupling $\lambda$.}
\label{penguin}
\end{figure}

Although many processes can be generated by the R-parity violating interactions, 
we focus on the LFV decays $\ell_m^- \to \ell_i^- \ell_j^- \ell_k^+$, where $m=\mu$ or $\tau$,  
in this subsection. As mentioned in the subsection {\bf 3.1}, the operator $\lambda L_3 L_I E_I^c$ 
generates the decays $\ell_m^- \to \ell_i^- \ell_j^- \ell_k^+$ at tree level when $\lambda \neq 0$. 
The other two operators in Eq.(\ref{rparity}), $\lambda'_{1,2}LQD^c$, also generate the similar decay processes
at one loop level through photon penguin diagrams shown in Fig. \ref{penguin}, 
but we found that the bounds on 
$\lambda'_{1,2}$ are stronger than that on $\lambda$ in the previous subsections. 
So we neglect contributions from $\lambda'_{1,2}$ operators to the decays
 $\ell_m^- \to \ell_i^- \ell_j^- \ell_k^+$.   
Moreover, flavor changing $Z$ boson decay $Z \to \ell^-_i \ell^+_j$ induced by the R-parity 
violating bilinear terms can contribute to  $\ell_m^- \to \ell_i^- \ell_j^- \ell_k^+$ processes. 
Since branching ratios of these decays are propotional to $\sin^2 \xi$, their effects are also negligible
\cite{bisset}. Besides these R-parity violating contributions, there are two other contributions to these 
processes by Higgs bosons. Since the charged leptons couple to the neutral Higgs bosons, 
these Yukawa interactions generate LFV processes at one loop level \cite{kolda}. 
However, these effects are enhanced only when $\tan \beta$ is large. 
So, we assume that these are negligible because $\tan \beta$ is small enough. 
Moreover, since there are three generations of both up and down type Higgs doublet in this model, 
LFV processes mediated by the neutral Higgs bosons are generated at tree level (Fig. \ref{penguin}).
However, the branching ratio of the $\mu \to eee$ from these effects is $BR \sim 10^{-16}$ 
when the neutral Higgs boson mass is $100 \GeV$ 
because of the smallness of the Yukawa couplings. So, these contributions can also be negligible 
compared to those from $\lambda LLE^c$ couplings unless $\lambda<10^{-3}$. 
Therefore, we can 
approximate that $\ell_m^- \to \ell_i^- \ell_j^- \ell_k^+$ processes are induced at tree level 
only by 
$\lambda$. In this approximation, the ratios of these processes are independent of $\lambda$, 
but depend on the mixing matrices $U_{eL(R)}$ which reflect the flavor structure of the 
model. Therefore we find some predictions of LFV decays $\ell_m^- \to \ell_i^- \ell_j^- \ell_k^+$ in our model.

From the branching ratio Eq.(\ref{mueeebr}), we can easily find the ratios of processes in the 
approximation that all scalar masses are equal:\footnote{From the conditions to suppress 
$\mu \to e+\gamma$ process from the scalar mass terms, slepton masses are required to be degenerated with mass differences of order $10^{-1}$\cite{kajiyama,kubo2}. }
\be
\frac{BR(\tau \to eee)}{BR(\tau \to \mu \mu \mu)}
&\simeq& \frac{4 \epsilon_{\mu}^2}{1 +\epsilon_{\mu}^2}=0.014, \\
\frac{BR(\tau \to \mu\mu e)}{BR(\tau \to \mu \mu \mu)}
&\simeq& \frac{1-\epsilon_{\mu}^2+2 \epsilon_e^2}{1+\epsilon_{\mu}^2-2 \epsilon_e^2}
=0.99, \\
\frac{BR(\mu \to eee)}{BR(\tau \to eee)}
&\simeq& \frac{\tau_{\mu}}{\tau_{\tau}}\epsilon_{\mu}^5 
\frac{\epsilon_e^2}{2 \epsilon_{\mu}^2+\epsilon_e^2}=0.0093,
\ee
where small parameters $\epsilon_{e,\mu}$ are given in Eq.(\ref{epsilon}) and $\tau_{\mu}(\tau_{\tau})$ stand for the lifetime of the $\mu (\tau)$ lepton. 
Also, $BR(\tau \to \mu \mu e)$ means $BR(\tau^- \to \mu^- \mu^- e^+)$, 
and similar for the other processes.
One can obtain the ratios of 
other combinations from the branching ratios listed below:
\be
BR(\mu \to eee)&\propto& \tau_{\mu} \epsilon_e^2 \left[ 
2(1-2 \epsilon_{\mu}^2)m^{-4}_{\tilde \ell 1L}
+\frac 12 m^{-4}_{\tilde \ell 3L}
-2(1-\epsilon^2_{\mu})m^{-2}_{\tilde \ell 1L}m^{-2}_{\tilde \ell 3L}\right], \\
BR(\tau \to eee)&\propto& \tau_{\tau}\left[ 
\epsilon^2_{\mu}\left( 1-\epsilon^2_{\mu}-4\epsilon^2_e\right)m^{-4}_{\tilde \ell 1L}
+\frac 12 \epsilon^2_e m^{-4}_{\tilde \ell 3L}\right],\\
BR(\tau \to \mu \mu \mu)&\propto&\tau_{\tau}\left[ 
\frac 14 \left(1+\epsilon^2_{\mu}-2 \epsilon^2_e-4 \epsilon^2_e \epsilon^2_{\mu} \right)
m^{-4}_{\tilde \ell 3L}
-2 \epsilon^2_e \epsilon^2_{\mu}m^{-2}_{\tilde \ell 1L}m^{-2}_{\tilde \ell 3L}\right],\\
BR(\tau \to ee\mu)&\propto&\tau_{\tau}\epsilon^2_e \left[ 
2 \epsilon^2_{\mu}m^{-4}_{\tilde \ell 1L}
+(\frac 12-\epsilon^2_{\mu})m^{-4}_{\tilde \ell 3L}
-2 \epsilon^2_{\mu}m^{-2}_{\tilde \ell 1L}m^{-2}_{\tilde \ell 3L}\right],\\
BR(\tau \to \mu \mu e)&\propto& \tau_{\tau}\frac 14 \left(1-\epsilon^2_{\mu}
+2 \epsilon^2_e-4  \epsilon^2_e \epsilon^2_{\mu} +\frac 14 \epsilon^4_{\mu}  \right)
m^{-4}_{\tilde \ell 3L},\\
BR(\tau \to \mu ee)&\propto&\tau_{\tau}\left[ 
\epsilon^2_e \epsilon^2_{\mu}m^{-4}_{\tilde \ell 1L}
+\frac 18\left( 1-2 \epsilon^2_{\mu}+6\epsilon^2_e \epsilon^2_{\mu}
+\frac 32 \epsilon^4_{\mu}\right)m^{-4}_{\tilde \ell 3L}  \right],\\
BR(\tau \to \mu e \mu)&\propto& \tau_{\tau}\left[ 
\frac 18 \left( 1-4 \epsilon^2_e+6 \epsilon^2_e \epsilon^2_{\mu}
-\frac 34 \epsilon^4_{\mu}\right)m^{-4}_{\tilde \ell 3L}
+2 \epsilon^2_e \epsilon^2_{\mu}m^{-2}_{\tilde \ell 1L}m^{-2}_{\tilde \ell 3L}\right],
\ee
where the common factor is not shown explicitly.


\section{Conclusion}

We have considered the properties of R-parity violating operators 
in a SUSY model with non-Abelian discrete $Q_6$ family symmetry. 
The family symmetry can reduce the number of parameters in the Yukawa 
sector, and explain the fermion masses and mixings between generations. 
It can also reduce the number of R-parity violating couplings 
and determine the form of those. Only three trilinear couplings are allowed, and 
the baryon number violating operators are forbidden by the symmetry in our model. 
We derived upper bounds on these couplings: $\lambda<O(10^{-2})$, 
$\lambda'_{1,2}<O(10^{-3})$, and obtained the predictions on the ratios of the LFV decays 
$\ell_m^- \to \ell_i^- \ell_j^- \ell_k^+$ which do not depend unknown parameters. 
The results reflect the properties of the family symmetry because these predictions contain 
the mixing matrices of the charged lepton sector which is written by masses of the charged leptons. 
Our predictions can be testable at future experiments because the superB factory \cite{kekb} or LHC 
\cite{tau3mu} will have the sensitivity $BR\sim 10^{-(8 - 9)}$ for LFV $\tau$ decays.

\vspace{0.5cm}
\noindent
{\large \bf Acknowledgments }\\
The author would like to thank the organizers for invitation to the conference. 
This work is supported by the ESF grant No. 6190 and postdoc
contract 01-JD/06.                                                                           

\end{document}